\documentclass[12pt]{article}
\usepackage{graphicx}
\newcommand{\arcsec}{^{\prime\prime}}

\newcommand{\arcmin}{^{\prime}}
\begin{document}
\title{Analysis of polarized microwave emission of Flare-Productive
Active Region 9415}
\author{$^{1}$V.M.Bogod\thanks{vbog@gao.spb.ru}, $^{2}$G.B.Gelfreikh,
$^{3}$F.Ch.Drago, $^{4}$V.P.Maximov,\\$^{5}$A.Nindos,
$^{1}$T.I.Kaltman, $^{6}$B.I.Ryabov,
 $^{2}$S.Kh.Tokhchukova}
 \date{}
\maketitle
\begin{center}
$^{1}$Special Astrophysical Observatory, Nizhnij Arkhyz, Russia
\end{center}

\begin{center}
$^{2}$Pulkovo Astronomical Observatory, St.Petersburg, Russia
\end{center}

\begin{center}
$^{3}$Institut of Solar-Terrestrial Physics, Irkutsk, Russia
\end{center}
\begin{center}
$^{4}$ Universita degli Studio di Firenzi, Florence, Italy
\end{center}

\begin{center}
$^{5}$University of Ioannina, Greece
\end{center}

\begin{center}
$^{6}$Ventspills International Radioastronomical Center, Riga,
Latvia
\end{center}

\begin{abstract}
The results of the microwave observations of the Sun made with
the RATAN-600 have shown the existence of many types of spectral
peculiarities in polarized emission of active regions, which
produce powerful flares. These phenomena happen at microwaves
and reflect inhomogeneous structure of magnetic field in
magnetospheres of flaring active regions in wide range of
heights above the photosphere. In this presentation we
demonstrate an analysis of the AR 9415 during all the period of
its passage across the solar disk. Results of the study point
out to existence of different scenarios of circular polarization
variations in the radio wave band. Here, we separated the
phenomenon of the cyclotron emission passage through the
quasi-transverse magnetic field (QT-region) and several effects
connected with  flare activity of active region.  New
observational data are presented and compared with the data of
several observatories: SSRT, NoRH, MDI SOHO, GOES and MEES. The
preliminary interpretation of the phenomena are given.
\end{abstract}

\section{Introduction}

The nature of flare-productive active regions (FPAR) on the Sun
is enriched by diversity of polarization emission features in
wide range of microwaves. It was found from detailed
spectral-polarization study that solar active regions
demonstrate various polarization peculiarities depending on
their flaring activity. The well-known phenomenon of circular
polarization inversion due to radio emission passage  through
quasi-transverse magnetic field region was detected and studied
in detaila in papers \cite{peter74}, \cite{kundu84},
\cite{gelf87}. The coupling mode theory created in the works
\cite{cohen60} and \cite{zhelez70} is the physical base for
interpretation of this phenomena in observations. Due to the
close geometrical binding to spatial height structure of
magnetic field (it must be located perpendicularly to the line
of sight), it is possible to reconstruct a vertical magnetic
field distribution up to several hundred Mm.

From the other side, the dynamic processes going on in active
regions frequently complicate the spectral and temporal behavior
of QT-effect. Therefore the main aim of this paper is to show
the main difference of QT-effect from other polarization
phenomena intrinsic to flare processes in AR.  As for the AR
9415, evolution of its circular polarization structure was very
complicated. Nevertheless, the evolution of polarization
inversion due to the QT-phenomenon was recorded clearly both on
time and on spectrum. At the same time the observed circular
polarization in sunspot associated sources caused by cyclotron
emission due to strong magnetic fields, changed greatly up to
multiple inversions. It points out to the influence of
multi-arch magnetic structure to transmitted polarized microwave
emission.

Recently, in the paper \cite{bogod03} it was shown that
polarizated microwave emission of FPAR's demonstrate some very
peculiar features, mostly in the wavelength range from 2 to 5
cm. Among them, a narrow-band short-wave (normally shorter than
3.5cm) polarization inversion and polarization brightening,
designated correspondingly as effects A1 and A2. Also multiple
spectral inversions inside of 10-20\% of frequency (named as
effect B), and a wide band feature of microwave "darkening" of
emission intensity \cite{tokh03}. All of these features were
registered in the AR 9415 both on the early preparatory stage
during several days before the powerful events and on the
eruptive stage of flare.

Several features in polarization inversions were detected in
early radio observations using the movement of AR along the disc
\cite{peter74}. Additional fine features were found using model
idea about geometry of quasi-transverse passage \cite{ryabov99}.
In particular, spectral-polarization behavior of sunspot
associated source emission during its multiple passage through
QT-regions was studied \cite{ryabov03}. From the other side, the
effects of multiple spectral and temporal polarization
inversions intrinsic to FPAR do not have correct interpretation
yet. Theoretically, other versions of multiple inversion besides
the QT-phenomena are possible. For example,  the following
mechanisms were considered by several authors: passage of
emission through regions with a high turbulence level, with high
electron density, and others \cite{bogodkot03}; inversion due to
the change of longitudinal magnetic field sign at the top of
coronal arch during the passage of AR along the solar disc
\cite{aliss84}, \cite{aliss93}; passage of cyclotron emission
trough current sheets \cite{gopalswamy94} and passage of region
with inverse gradient of temperature with height
\cite{zlotnik01}, \cite{kaltman03}, etc.

\section {Observations}

\subsection {RATAN-600}

For the research we have used opportunities of the
radiotelescope RATAN-600 (located in Northern Caucasus, Russia),
which has optimum parameters for the research of preflare plasma
in active regions. For the current task it was the  most
important the combination of an instant spectrum in a wide
frequencies range (1.7cm - 8cm) with a good spectral resolution
($5 - 7\% $), high accuracy of polarization measurements ($<
0.05\% $), and high brightness temperature sensitivity ($<10$K).
RATAN-600 has a moderate spatial resolution in a horizontal
plane ($<15\arcsec $ at 2cm) and relatively low resolution in a
vertical plane ($\sim15\arcmin$ at 2cm), which varies
proportionally to the wavelength. Observations were carried out
daily from about 7:00 UT to 11:00 UT with a time cadence of 8
minutes. The obtained one-dimensional images ("scans") of the
Sun are published at:$http://www.sao.ru/hq/sun/$
\cite{bogod03b}.

\subsection{SSRT}

Siberian Solar Radio Telescope (SSRT)  is a heliograph
consisting of 256 antennas and working at the wavelength of 5.2
cm. It obtains two-dimensional radio maps of the sun with
spatial resolution about $20\arcsec$ each 3-4 minutes. We have
used observations presented in Internet:
\\{http://$ftp://iszf.irk.ru/pub/ssrt\_data/fits/$.

\subsection {NRH}

The 2D observations made with Nobeyama radioheliograph (NRH) at
the wavelength 1.76cm with spatial resolution $10\arcsec\times
10\arcsec$ in intensity and circular polarization were taken
from the site: \\
$ftp://solar.nro.nao.ac.jp/pub/norp/data/daily/$. For the
comparison with RATAN-600 data the 2D maps were convolved with
the RATAN-600 beam at wavelength 1.83cm.

\subsection {SOHO MDI and GOES}

The satellite data were used for obtaining of information about
the flare class (GOES) and magnetic field structure on the
photosphere level (SOHO MDI). The magnetograph data of SOHO MDI
in FITS format with 96 minute cadence were obtained from the
site: $http://soi.stanford.edu/magnetic/mag/$ and GOES data from
$http://www.sec.noaa.gov$. In the necessary cases a convolution
with the different wavelength beams of RATAN-600 was performed.
Up to now the satellite data are the good base for associating
the radio data.

\subsection {Mees data}

The data of this observatory were used for quick determination
of magnetic field configuration of the AR 9415. They were taken
from the site\\ http://www.solar.ifa.hawaii.edu/ARMaps/.

\section{Total information about the active region AR 9415}

Let us consider development of our AR in period from April 3
till April 15 of 2001. Magnetic configuration of the region was
rather complicated. According to the data obtained at the
observatory Mees, it passed through following stages of
development: $\beta$ (3-4.04), $\beta-\gamma$ (5-7.04),
$\beta-\gamma-\delta$ (8-13.04), $\beta-\gamma$ (14.04), and
again  $\beta$ (15.04). 69 flares were registered in the region,
including  6 X-ray bursts of I level and 5 of O. We have used
the RATAN-600 to study spectral polarization structure of the AR
and radio maps obtained with the SSRT to get its two-dimensional
structure at wavelength of 5.2cm. Space observations from GOES
and  SOHO MDI were used to get information on parameters of
flares and the structure of the photospheric magnetic fields.
The Table 1 summarizes information on evolution AR9415 found
from all the above instruments during the  period under study.

\begin{table}
\caption{ }
\begin{tabular}
{|p{28pt}|p{42pt}|p{79pt}|p{84pt}|p{92pt}|} \hline
\par Date & \textbf{MDI} \par\textbf{SOHO}, \par \textbf{MEES}
\par & \textbf{GOES}\par & \textbf{RATAN-600}\par & \textbf{SSRT}\par\\

\hline 3.04.& $\beta$& X1.2(3h57m)& RH,\par monotonous\par
spectrum& RH,single source\\

\hline 4.04& $\beta$ & M1.6(10h27m)&RH,\par monotonous\par
spectrum & Bypolar source\par RH,W-side,strong\par LH, E-side,
weak\\

\hline 5.04& $\beta-\gamma$& M5.1(17h25m)& RH,short-wave\par
increase& Single,RH,\par W-side,\\

\hline 6.04& $\beta-\gamma$& X5.6(19h21m)& RH,short-wave\par
increase& Single,LH,center \\

\hline 7.04& $\beta-\gamma$& Stable&LH,short-wave\par increase&
Bypolar,\par LH, center, strong\par RH,E-side,weak \\

\hline 8.04& $\beta-\gamma-\delta$& Stable& LH,short-wave\par
increase& LH,center,strong \par RH,E-side,weak\\

\hline 9.04& $\beta-\gamma-\delta$ &M7.9(15h34m)& No data& LH,
center, strong\par RH,E-side,strong\\

\hline 10.04& $\beta-\gamma-\delta$&X2.3(5h26m)&
LH,short-wave\par increase& LH,center,strong
\par RH,E-side,strong \\

\hline 11.04& $\beta-\gamma-\delta$&M2.3(13h26m)&
LH,short-wave\par increase& LH,center,weak
\par RH,E-side,strong \\

\hline 12.04& $\beta-\gamma-\delta$&X2.0(3h04m)\par
M1.3(10h28m)& LH,short-wave\par increase& LH,center,weak\par
RH,E-side,strong \\

\hline 13.04& $\beta-\gamma-\delta$&Stable& RLR inversion&
Single RH, strong
\\

\hline 14.04& $\beta-\gamma$&M1(18h11m)& RLR inversion&Single
RH, strong
 \\

 \hline 15.04& $\beta$&X14.4(13h50m)& RLR inversion&No data
 \\
\hline
\end{tabular}
\end{table}

\section{Three periods of evolution of the AR 9415}
Three periods of powerful flaring activity one can follow in the
AR 9415 using GOES data on Figure~\ref{fig3}) and the Table I.
The first period was from April 3 to 6,  when weak activity of
the AR 9415 took place on the ground of powerful activity of the
preceding AR 9393. The second period was registered from the
April 7 to 14 and the third from 12 to 14. Each period was
beginning from low activity and was finished by generation of
powerful flares of the level M or X.

It is interesting to note that does not seem any correlation
between microwave polarization spectra (observations made with
the RATAN-600 and  SSRT) and variations in magnetic structure at
the photospheric levels (MDI SOHO and Mees data). The latter
showed development to more complicated magnetic structure, from
$\beta$ (April 3) to $\beta-\gamma-\delta$ configuration (from
April 8 to 13).

\subsection{Period April 3-6 }
At the Figure~\ref{fig1} the comparison of the radio and optical
data during April 3 is presented.  Figure~\ref{fig1}a shows the
location of the AR 9415 on  SOHO MDI magnetogram (to the left)
and 2D map from SSRT at wavelength 5.2cm (to the right).  It is
worth mentioning a very complicated $\beta$ configuration of the
magnetic field and dominated one sign RH polarized source in the
region. On the Figure~\ref{fig1}b the scans of the Sun made with
the RATAN-600 are shown at several cm wavelengths. At the top it
is shown  the scans in intensity and below the scan in circular
polarization. One can see that the source of AR 9415 at E limb
consist of two components A and B. East source A is located on
the distance of $40 - 45\arcsec$ (it corresponds to the height
about 30 thousand km above photosphere) and demonstrates the
inversion of circular polarization from short wave to longer.
The other more power source B is located near the limb and has
the RH polarized emission.  On the Figure~\ref{fig1}c at the top
the spectrum of polarized emission of the Eastern source is
shown. This has an inversion of the sign of polarization around
the frequency of 11 GHz. At the lower figure we see the spectrum
of the main source, one typical for a cyclotron sunspot
connected emission. One can suppose that the presence of the
higher East component with nonthermal spectrum has  something to
do with high flaring activity of the AR 9415.

On the Figure~\ref{fig2} the comparison of radio and optical
data is presented for the April 5 and 6. The magnetic field map
(left column - to the top picture for April 5 and the down for
April 6) according  SOHO MDI demonstrates the stable structure
with small changes. On the contrary the radio map at 5.2cm
obtained on the SSRT (second column: up for April 5, down one
for April 6) shows polarization inversion from the RH to LH.
This is probably due to the effect of wave propagation through
the region of QT magnetic field.

At the right the RATAN-600 scans  at a number of cm wavelength
are presented. Let us note that the sign of polarization
registered at SSRT at $\lambda$=5.2cm (SSRT) agrees with that at
$\lambda$=4.93cm observed at RATAN-600. However, already at
$\lambda$=6.06cm the inversion of polarization in QT region
takes place. One should also notice the increase of the
polarized emission at short waves (2.32cm~-~1.92cm). This kind
of event is often connected with appearance of new magnetic flux
in the existing AR \cite{bogod03}.

According to the observations made with GOES (see
Figure~\ref{fig3}) the level of flare activity of the AR 9415 in
this period was not too high and was observed at the background
of the preceding  more active region AR 9393.

\begin{figure}
\centerline{\includegraphics[totalheight=13cm]{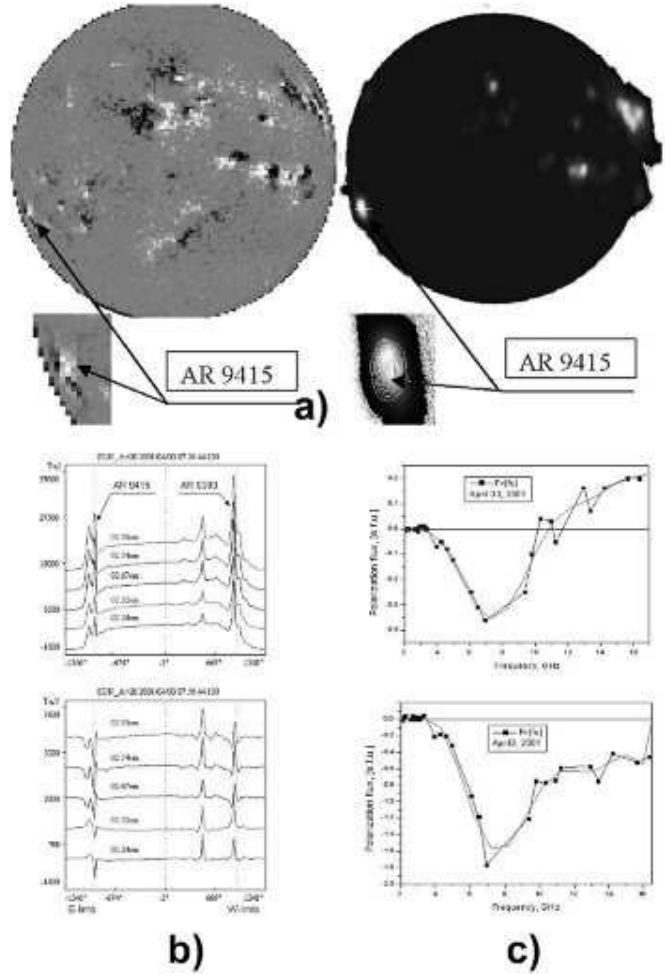}}
 \caption{The rising of AR 9415 on the E-limb. a) On the left
 the solar disc magnetic field map according MDI SOHO data;
 on the right - the radio map at wavelength 5.2cm made with SSRT,
  below - the polarized emission map for  AR 9415.
 a)One-dimensional radio scans of AR 9415 made with RATAN-600 during April 3,
2001.
 Up - the intensity scans at different waves moved in vertical axis.
 Below-the circular polarization scans ( righthanded polarization is negative).
 AR9415 consist of two radio sources A (above the limb) and B (more intensive).
 c) The spectra of polarization emission: up - for source A,  and down - for
source B. The smoothed curves with polynomial fit are shown by
dotted lines.} \label{fig1}
\end{figure}

\begin{figure}
\centerline{\includegraphics[totalheight=6.5cm]{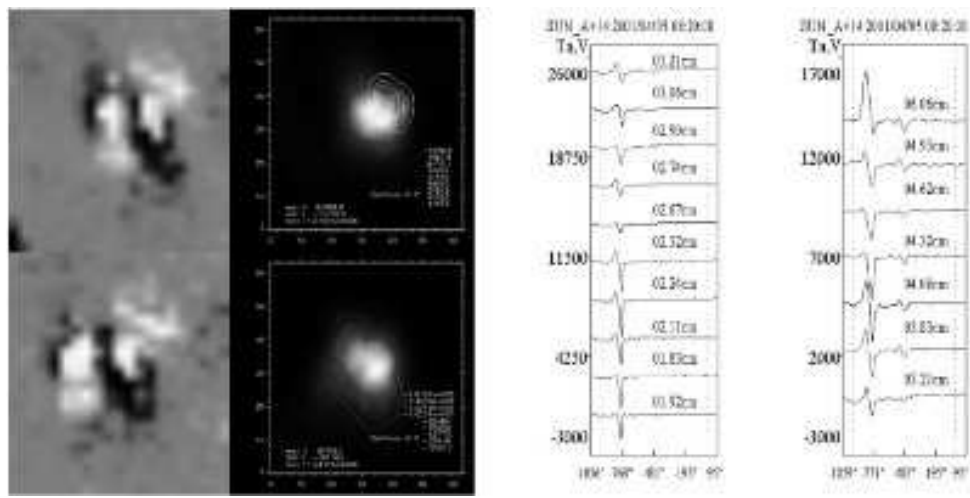}}
 \caption{Comparison of photosphere magnetic field data and radio polarization data
for AR 9415. On the left - MDI SOHO data, up - MDI map for April
5, down - MDI map for April 6. In the second column: up- the
radio map at 5.2cm for April 5, down - the same for April 6. One
can see the QT inversion at this wave. To the right, the
one-dimensional RATAN-600 polarization emission scans at several
wavelengths for April 4, 5 and 6 are presented. One can see the
short wave brightening of polarization emission for April 5 and
6. The polarization inversion at longer wavelength 6.06cm is
recorded also.} \label{fig2}
\end{figure}

\begin{figure}
\centerline{\includegraphics[totalheight=4.5cm]{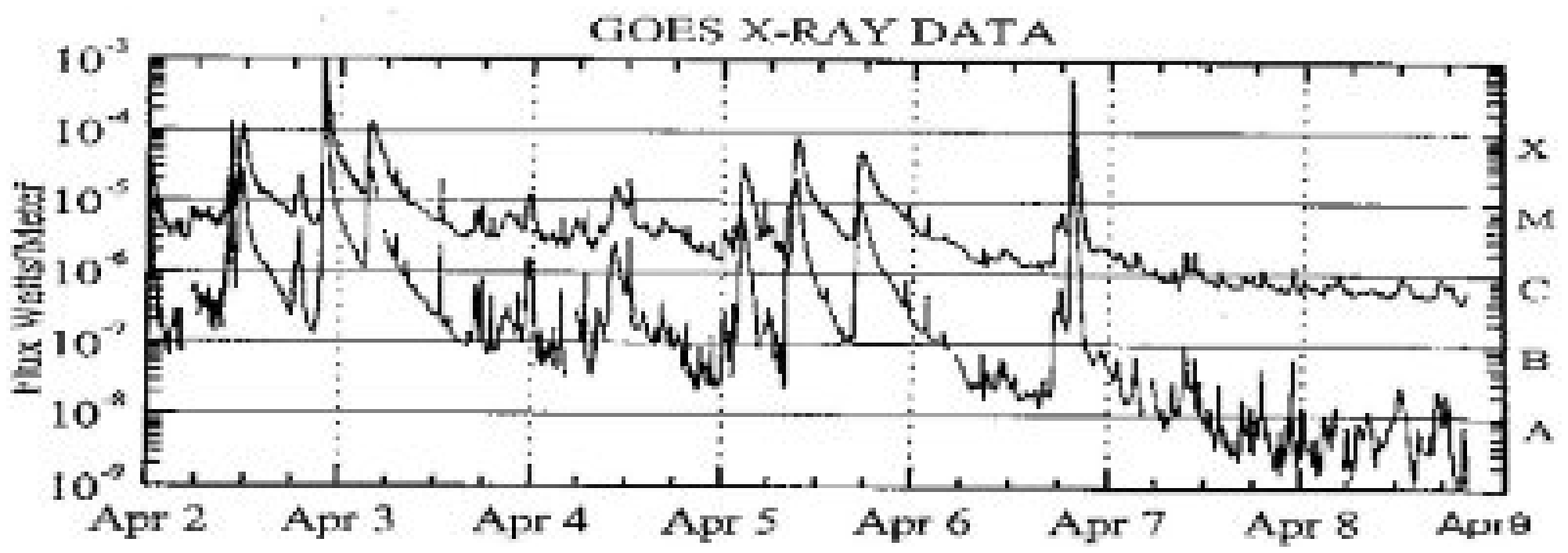}}
 \caption{Soft X-ray GOES data during interval from  April 2 to April 16, 2001.}
 \label{fig3}
\end{figure}

\subsection{Spectra and polarization of the AR9415 in the period of April 6~-~7}
According to the RATAN-600 observations the sharp inversion of
polarization took place in wide wavelength range from 1.83cm to
3.06cm (see Figure~\ref{fig4}). The variations in the structure
of the magnetic field at the photosphere according to the
optical magnetograph of  MDI SOHO were also observed but they
did not seem so significant. Observations made with the SSRT
showed appearance of a new radio source at the East of  the AR
9415.

For more detailed comparison of the radio and optical
observations the convolution of optical magnetogram with diagram
pattern of the RATAN-600 was made for wavelength
$\lambda$=2.24cm (17$\arcsec\times15\arcmin$ size). The results
are shown on the Figure~\ref{fig5} for several moments of
observations on April~6~and~7. At this Figure the central part
of optical magnetogram is shown which is registered by the RATAN
scan of the solar disk. In general one can see that
one-dimensional structure of the photospheric magnetic field was
stable during these two days period (see Figure~\ref{fig5}).

On the Figure~\ref{fig6} the comparison of the radio scans at
the wavelength of 2.24cm and integrated to one-dimensional scale
magnetograms of MDI SOHO for two times and the two dates, April
6 and 7. One can see that at the radio waves total reversal of
the sign of the circular polarization while at the photospheric
level no significant variation of the magnetic field structure
was registered. The sign of polarization in the radio wave range
become correspond to excess of  the extraordinary mode. Some
oscillations of the optical and radio magnetograms in their
related position are also observed in the range of tens of
arcsecs. In the time interval between the presented data for the
6 and 7 April a powerful flare happened on (class O5.6) at 19h
21m UT on April 6.

\begin{figure}
\centerline{\includegraphics[totalheight=5cm]{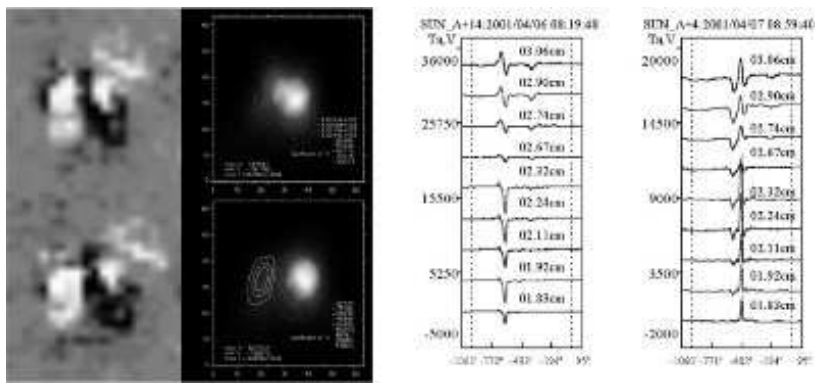}}
 \caption{Comparison of photosphere magnetic field data and radio polarization data
for AR 9415. On the left - MDI SOHO data, up - MDI map for April
6, down - MDI map for April 7. In the second column: up - the
radio map at 5.2cm for April 6, down - the same for April 7. To
be note the stable magnetic field structure on the photosphere
for both days. To the right, the one-dimensional RATAN-600
polarization emission scans at several wavelengths for April 6
and 7 are presented. One can see the wide band  of polarization
inversion was occurred between two days.} \label{fig4}
\end{figure}

\begin{figure}
\centerline{\includegraphics[height=7cm]{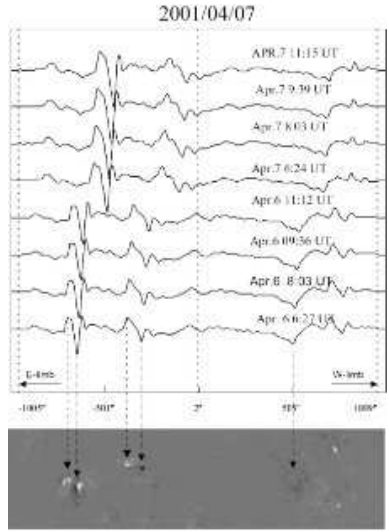}}
 \caption{One dimensional magnetic field scans, made in
 convolution of 2D map of MDI SOHO with one-dimensional RATAN-600 beam.
 One can see the stability of photosphere magnetic field structure
 during two days  for several time moments. The AR 9415 is located
 at east part of the disc. }
 \label{fig5}
\end{figure}

\begin{figure}
\centerline{\includegraphics[totalheight=7cm]{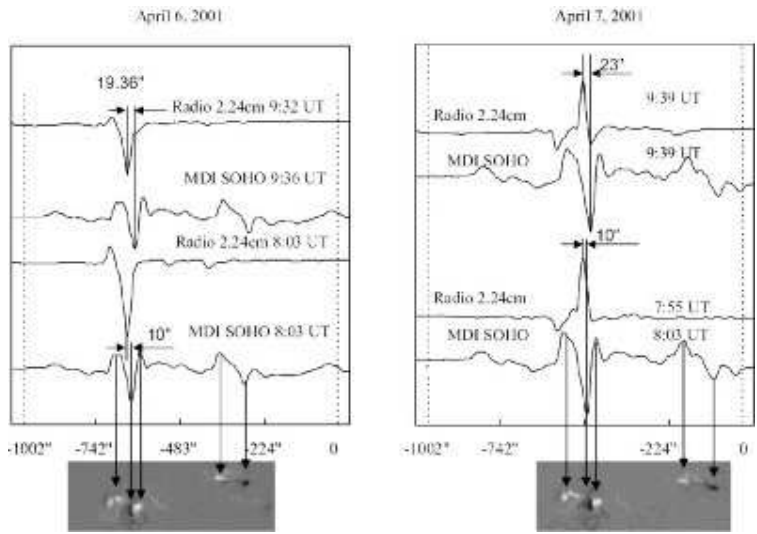}}
 \caption{Comparison of 1D radio scans at wavelength 2.24cm and 1D
 MDI SOHO,  for two time moments during April~6~and~7. This
 picture demonstrate the polarization inversion in radio with
 stable structure in optics.}
  \label{fig6}
\end{figure}

\begin{figure}
\centerline{\includegraphics[totalheight=6cm]{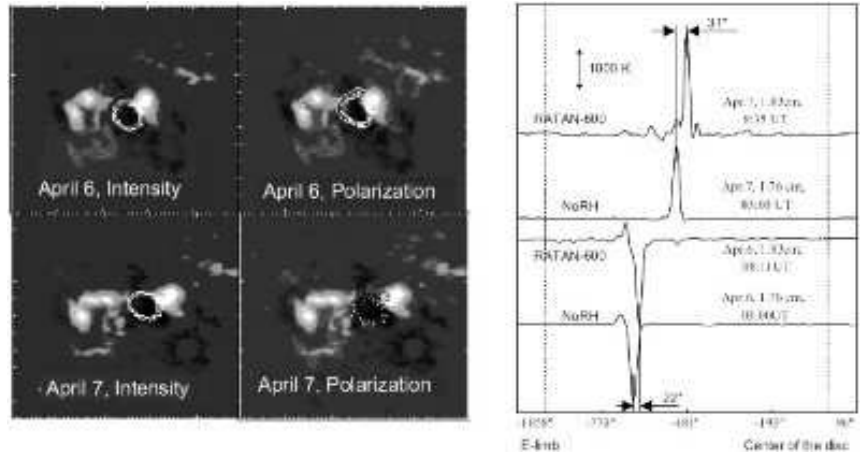}}
 \caption{Comparison the radio data for two instruments Nobeyama
 radioheliograph at 1.7cm and RATAN-600 at 1.83cm for two days
 April~6~and~7. To the left- the comparison of 2D data. To the
 right- the comparison of 1D data with using convolution
 procedure. The radio data from two different telescopes confirm
 each other. }
 \label{fig7}
\end{figure}

\subsection{Polarization spectra during the period April 8-12 \\ and their
peculiarities}

On the Figure~\ref{fig8} the spectral characteristics of AR
during from April 4 till April 15, 2001 are  compared with 2D
maps from  SSRT at 5.2cm. On the left  spectra of intensity
flux, on the center - spectra of the polarized emission
(righthanded polarization points downwards) for the central
intense radio source B are presented. On the right radio images
at 5.2cm in intensity and circular polarization are shown. Here
dotted contours correspond to the lefthanded polarization, and
solid line contours correspond to the righthanded one.

 It is interesting to note, that  a wide range polarization inversion
which occurred from April~6~to~7 did not change the  absolute
character of a spectrum. The enhancement in a short-wave part of
a spectrum preserved  on  April 7,  although  the sign of
polarization  turned to opposite (see Figure~\ref{fig8}). The
period from April 7 till April 11 was characterized by a steady
growth of  left polarization spectrum on short waves. This
process lasted up to April 12 and was accompanied by intense
X-ray flares, as seen on the Figure~\ref{fig3} and Table 1.
Figure~\ref{fig8} demonstrates that  SSRT registered the
appearance of new polarized source from the east side of the AR.
Its intensity was monotonously rising and became dominant in the
AR to the end of the period.

\begin{figure}
\centerline{\includegraphics[totalheight=14cm]{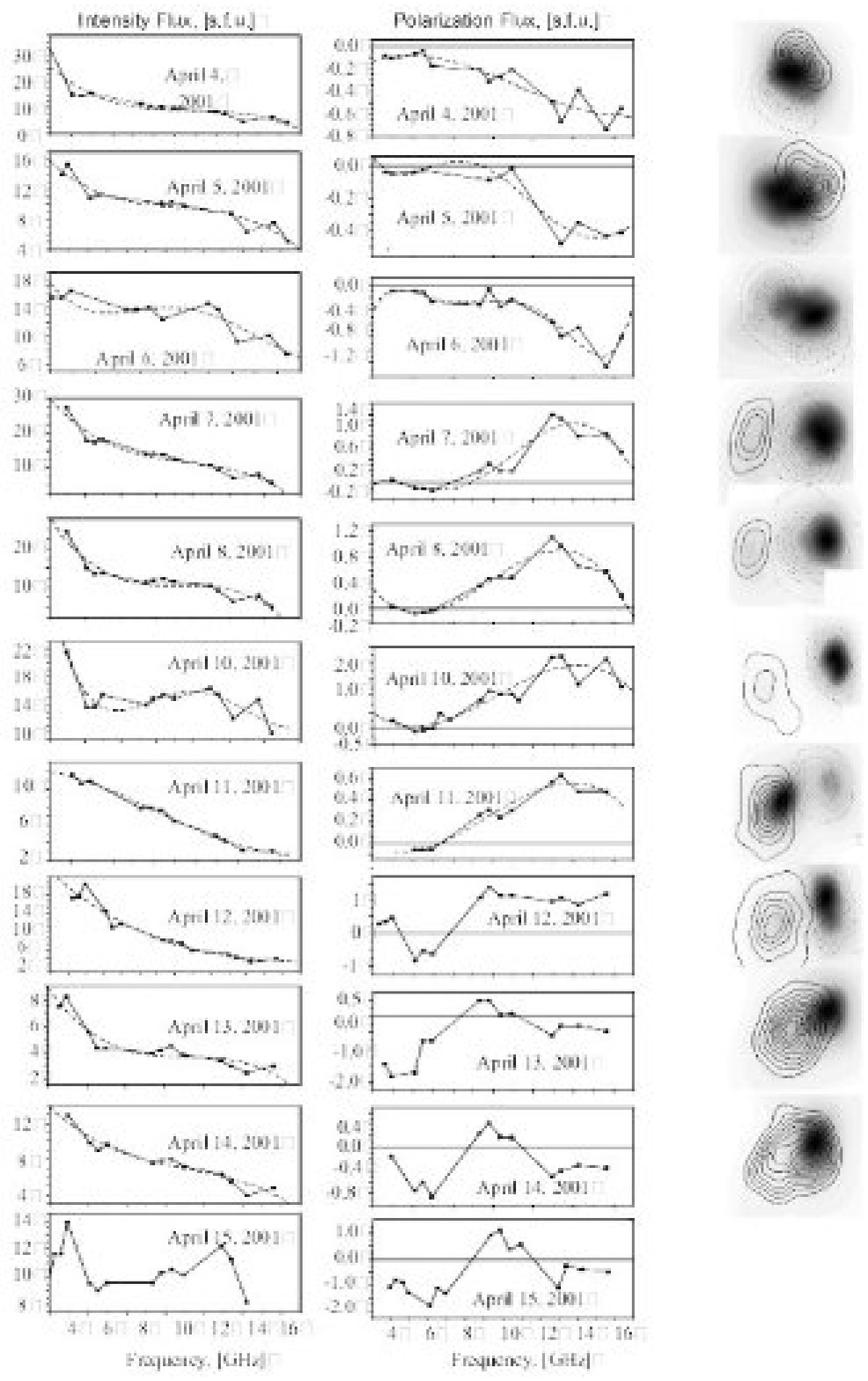}}
\caption{Comparison of RATAN-600 full flux spectra for intensity
(to the left) and for circular polarization (in the centrum, to
right- polarization is negative) with 2-D structure at 5.2cm.
SSRT data are shown on the right column. Polarization maps (
right polarization - solid line, left- dotted line) overlayed on
intensity maps (the brightness is shown black color)).
 } \label{fig8}
\end{figure}

\begin{figure}
\includegraphics[totalheight=10cm]{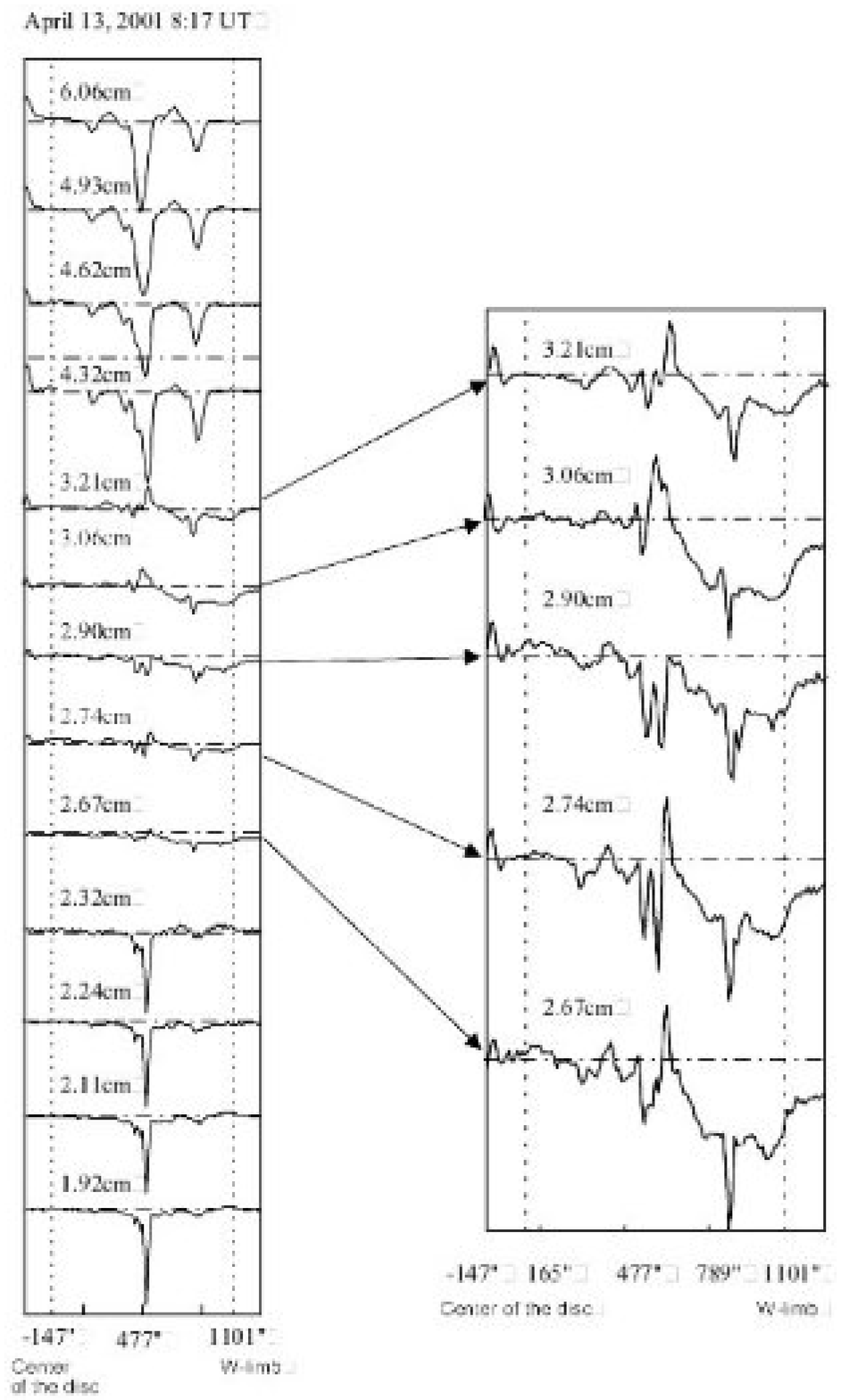} \hfill
\includegraphics[totalheight=12cm]{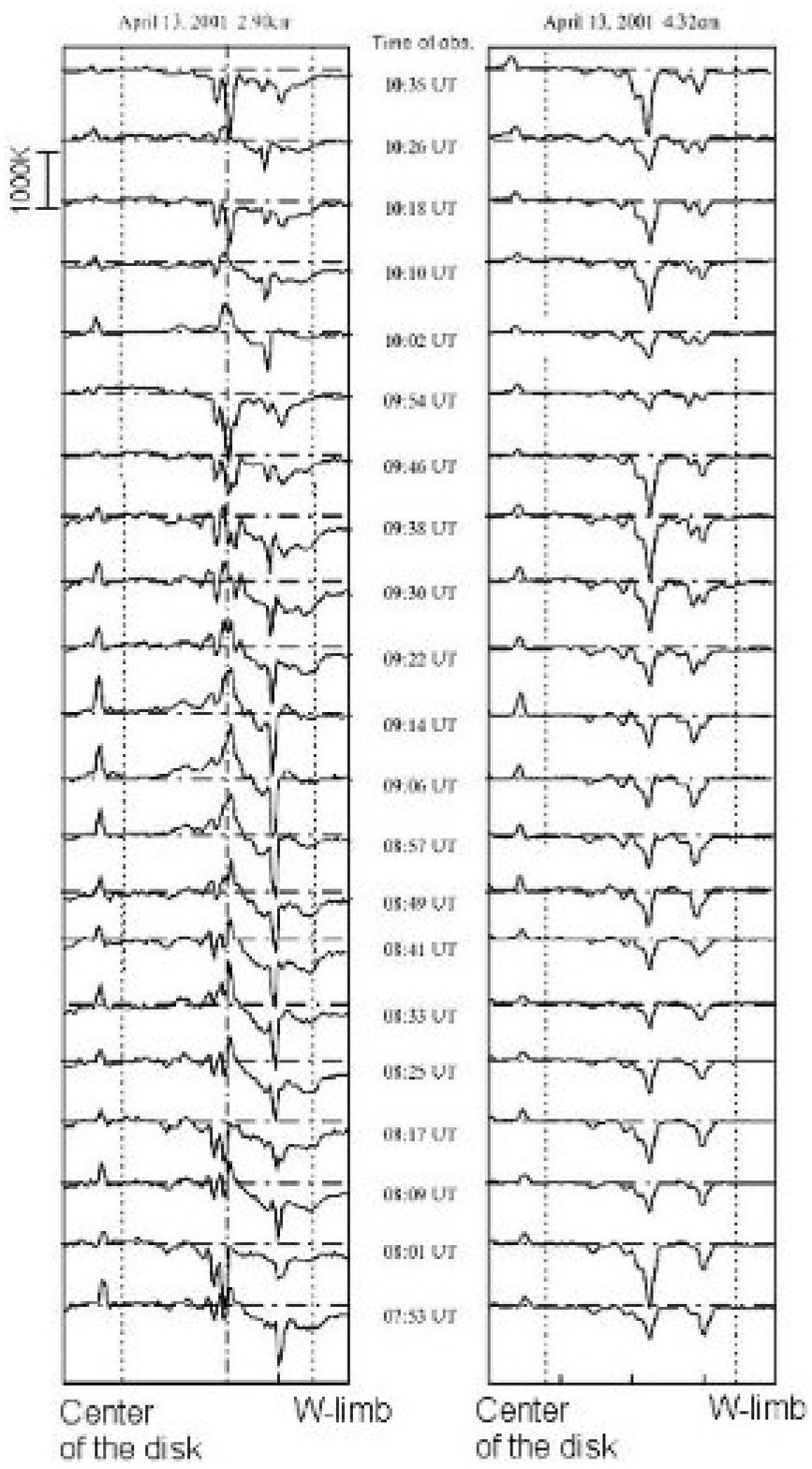}
\\
\parbox[t]{0.47\textwidth}{\caption{AR 9415 microwave emission polarization
structure for April 13, 8:17UT. One can see the spectral
peculiarity in the range 2.67cm - 3.2cm, where degree of
polarization get weaker, the polarization structure sharply
changing with wavelength. This is illustration of so named
Frequency Domain of Modes Coupling (FDMC).}\label{fig9}} \hfill
\parbox[t]{0.47\textwidth}{\caption{Evolution of AR 9415 polarization
structure during April 13, from 7:53 UT to 10:35 UT.  The
wavelength 2.90cm is located inside the FDMC, and 4.32cm is
outside the FDMC. Note multiple polarization inversions at
2.90cm on time. This phenomenon was registered in AR 9415 3 days
before a big flare X14.4 on April 15, 13h50m UT.}\label{fig10}}
\end{figure}

\subsection {Features of polarization spectra  during \\ April 13 - 15}

On April 12 - 13 the growth of a spectrum of the polarization
flux on short waves has terminated. It is shown on the
Figure~\ref{fig3}, that flare activity on April 13 and 14 was
considerably low. This reorganization has resulted in appearance
of a new form of spectrum (Figure~\ref{fig9}).

During this period on short wave range (1.83cm - 2.24cm) and on
long waves (longer than 3.21cm) the sign of circular
polarization has changed again to the righthanded. As seen on
Figures~\ref{fig9} and \ref{fig10}, in a spectrum of polarized
emission the intermediate frequency domain was formed, in which
narrow (point) radio sources with numerous inversions of a
circular polarization sign both on frequency and on time have
appeared. In the paper \cite{bogod03} the domain was named as a
"Frequency Domain of o- and e- Modes Coupling" (FDMC), because
of multiple inversions of circular polarization on time and on
frequency. Outside  this frequency domain the polarization
degree of the AR 9415 has decreased to small values $(<0.03\%)$
(Figure~\ref{fig9}), and in the channels of intensity
(Figure~\ref{fig11}) the "darkening effect" \cite{tokh03} has
developed . As well as for a case with known Bastille day flare,
this process has resulted in a powerful X14 flare at 13h 50m UT
on April 15 (Figure~\ref{fig3}).

\begin {figure}
\centerline{\includegraphics[totalheight=4.8cm]{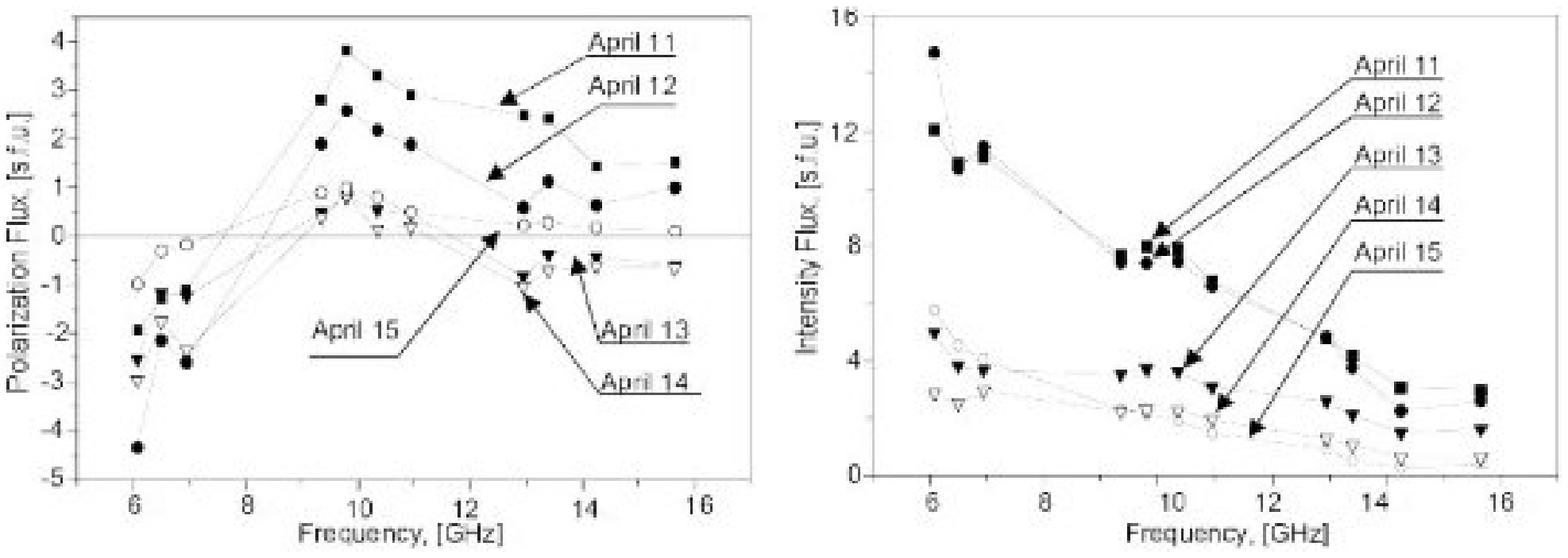}}
 \caption {The "darkening" effect before the O14.0 flare at 13h~50m~UT on April~15.
 To the right the intensity flux is decreased from day to day. To the left
 the polarization emission  also is decreased with forming the band with
 weak polarization (FDMC)}
\label{fig11} \end{figure}

\section{Discussion}

A variety of magnetic structures in solar AR atmosphere results
in large variety of effects in microwave polarized emission, its
variability on spectrum, on time, depending on flare activity
etc.

 It is well known, that the intensity and topology of coronal magnetic fields
 can influence the distribution and final polarization of radioemission from
 the underlying sources. The inversion of  polarization sign occurs,
 when the emission on its way to the observer crosses a region, where the
longitudinal component of an external magnetic field changes its
direction:  here the LH circular polarization of a radiosource
becomes RH and vice versa. The inversion of polarization can
also occur, if the radiation on different frequencies arises at
different locations. As a result of different propagations
through the area of inversion some parts of the source may
undergo polarization inversion, while other parts do not show
any polarization change.

Flare productive active regions demonstrate sharp changes of the
sign and magnitude of polarization at the early preliminary
phase of a flare as well as on its eruptive stage. It is widely
accepted \cite{aliss94}, that active processes on the Sun are
obliged to the release of energy  confined in magnetic fields.
However, at the early preparatory phase of a powerful solar
activity these processes take place on  much smaller level of an
energy-release. A question of places of localization of a
primary energy-release \cite{bastian96}  is not clear until now:
where, at what heights, and how the primary energy-release is
happening and the storage of energy in the magnetosphere of an
active region occurs, which further results in the instability
and realization of an active flare phase under the various
possible scenarios \cite{vrsnak03}. It was unexpected the
detection of fine structure of polarized radio emission spectra
of flare productive active regions at early stages, several days
prior to powerful flares \cite{bogod03}. It was shown, that such
an initial heating occurs at low corona levels and can be
registered with reflecting radio telescopes with high flux
sensitivity in the range of 2cm~-~5cm and detailed spectral -
polarization analysis. Changes in a spectrum of polarized
emission, inversions of polarization on frequency and on time,
practically, always take place at the early stages of flare
processes. The registered polarized emission from active region
at this stage can reach very small values from several hundredth
of s.f.u. up to several s.f.u. And the degree of circular
polarization can drop to some fractions of  \% . The analysis of
the emission of the FPAR 9415 gives a number of
 new facts on evolution of active region, including  several
characteristic properties marked in a radio range. Detailed
interpretation and modeling analysis will be stated in
subsequent papers. Here we want to state the most probable
versions of interpretation of the registered effects.

\subsection{The rise of a new magnetic flux}

The  new magnetic flux  rise is accompanied with an inversion
and brightening of polarized emission in short wave part of
centimetric range. The more probable hypothesis  assumes the
appearance of small islands of new rising magnetic flux with
strength of 1500 - 2000~G and with sizes much smaller than the
beam width $\theta_{x}~\theta_{y}$ at the level of low corona.
This rising magnetic flux could be of a north or south polarity
and could annihilate with an old (and more weak) magnetic flux
with heterogeneous structure.  Due to the high flux sensitivity
of RATAN-600 (since a big effective area of antenna, high
accuracy of polarization measurements, parallel multi wavelength
analysis) a small islands of a new magnetic field can be
registered at short wavelengths as a cyclotron emission. Let us
make some estimations of sizes of rising magnetic flux islands
for the case when the emission is fully polarized and it is
determined by emission of extra-ordinary wave in the area of
high coronal temperatures.

Taking into account the sensitivity in scan procedure with the
fan beam of RATAN-600 at wavelength  2.24cm with beam sizes
$\theta_{x}\theta_{y}= 17\arcsec\times 15\arcmin =
15300{\arcsec}^{2}$ for an island with sizes
$\theta_{xi}\cdot\theta_{yi}$ we can write:
\begin{equation}
 \theta_{xi} \cdot \theta_{yi} = \frac{{\theta_{x} \cdot
\theta_{y} \cdot 2\sqrt {2} \cdot T_{n\odot}} }{{T_{B} \cdot
\sqrt {\Delta f \cdot \tau} } }
\end{equation}
 where $T_{n\odot} $ is a noise temperature of the system, including instrument
noises $T_{n}$ and quiet Sun antenna temperature
$T_{\odot}$=$\alpha$$\cdot$$T_{B}$, approximately equaled 12000,
where $\cdot T_{B}$ is brightness temperature of island, and
$\alpha\simeq 2$.

With a bandwidth $\Delta f = 500$MHz, time constant
$\tau=0.2sec$, we obtain a size of magnetic island $\theta_{xi}
\cdot \theta_{yi}=5\arcsec$ for temperature $T = 10^{4}$K,
$\theta_{xi}\cdot\theta_{yi}=0.5\arcsec$ for $T = 10^{5}$K and
$\theta_{xi}\cdot\theta_{yi}=0.05\arcsec$ for $T = 10^{6}$K.
Thus, high sensitivity is very important even for observations
with moderate resolution of narrow and bright polarized details
on the Sun.

In case of the polarity of new magnetic field coincides with the
polarity of old magnetic field we register the rise of polarized
flux at short wavelengths. In opposite case the short wavelength
polarization inversion is observed. Thus from consideration for
example of Figure~\ref{fig2}, it follows that in the range of
1.92cm - 2.32cm we registered the rise of a new magnetic loop
with a field strength of 1950 G at the base level  and 1540 G at
the top of the loop. The predominance of the extraordinary mode
of emission on the 3-rd harmonic of gyrofrequency  was assumed
in these estimations.

 The short wavelength brightening in spectrum of polarized emission
 of both signs was registered till April, 12 (Figure~\ref{fig2}).
This period was accompanied with strong flares (see Table 1,
Figure~\ref{fig3} and \ref{fig8}) which argues for hypothesis of
 annihilation of a newly emerged magnetic flux with an old (and more weak)
magnetic
 fields. Since the beam width of RATAN-600 at these wavelengths
 is approximately $17\arcsec \times 15 \arcmin$ and covers a significant
  area of active region it
includes fine structured details of magnetic fields of both
polarities. The convolution of the polarization beam of the
radio telescope with an actual structure of radio brightness
distribution is the difference of integrals over the right
polarization and the left polarization. With assumption of
cyclotron mechanisms of emission this convolution corresponds to
the difference of magnetic fluxes at heights of radio emission
of centimeter band. So, the observed structure of polarized
emission is often complicated. The use of spectral-polarization
analysis allow to separate the radio sources based on frequency
spectra.

The discontinuance of short wavelength brightening in AR 9415 in
period after April, 12  was accompanied with a  cessation of
activity and with  preparation of a new scenario of flare energy
release (see below for period of April~13~-~15).

 \subsection { Recognized regularities of polarization inversion in AR
9415}

In this Section we try to retrieve and analyze the spectra
polarization  changes originated strictly from the
QT-propagation of microwaves through the solar corona on the way
to observer.

 \subsubsection {Analysis of QT-propagation}

    Let us retrieve the well-known effects of QT-propagation
    from the challenging history of the spectra polarization changes
    in AR 9415 in the time period of April 3 - 15, 2000. For the sake
    of simplicity beginning with the AR 9415 at the central solar meridian
    on April 9, we will proceed to the limb positions. 3 microwave sources
    clearly seen at short cm wavelengths can be associated with the dominated
    sunspots of N-S-N magnetic polarity as A, B, and C sources of R-L-R sense
    of circular polarization correspondingly Figure~\ref{fig13}. The above
correspondence
     is evidenced by the RATAN-600 V scans Figure~\ref{fig6} (the LH circular
     polarization is positive and the RH is negative throughout
     the scans). It implies the prevalence of extraordinary mode radiation.
\begin{figure}
\includegraphics[totalheight=5.8cm]{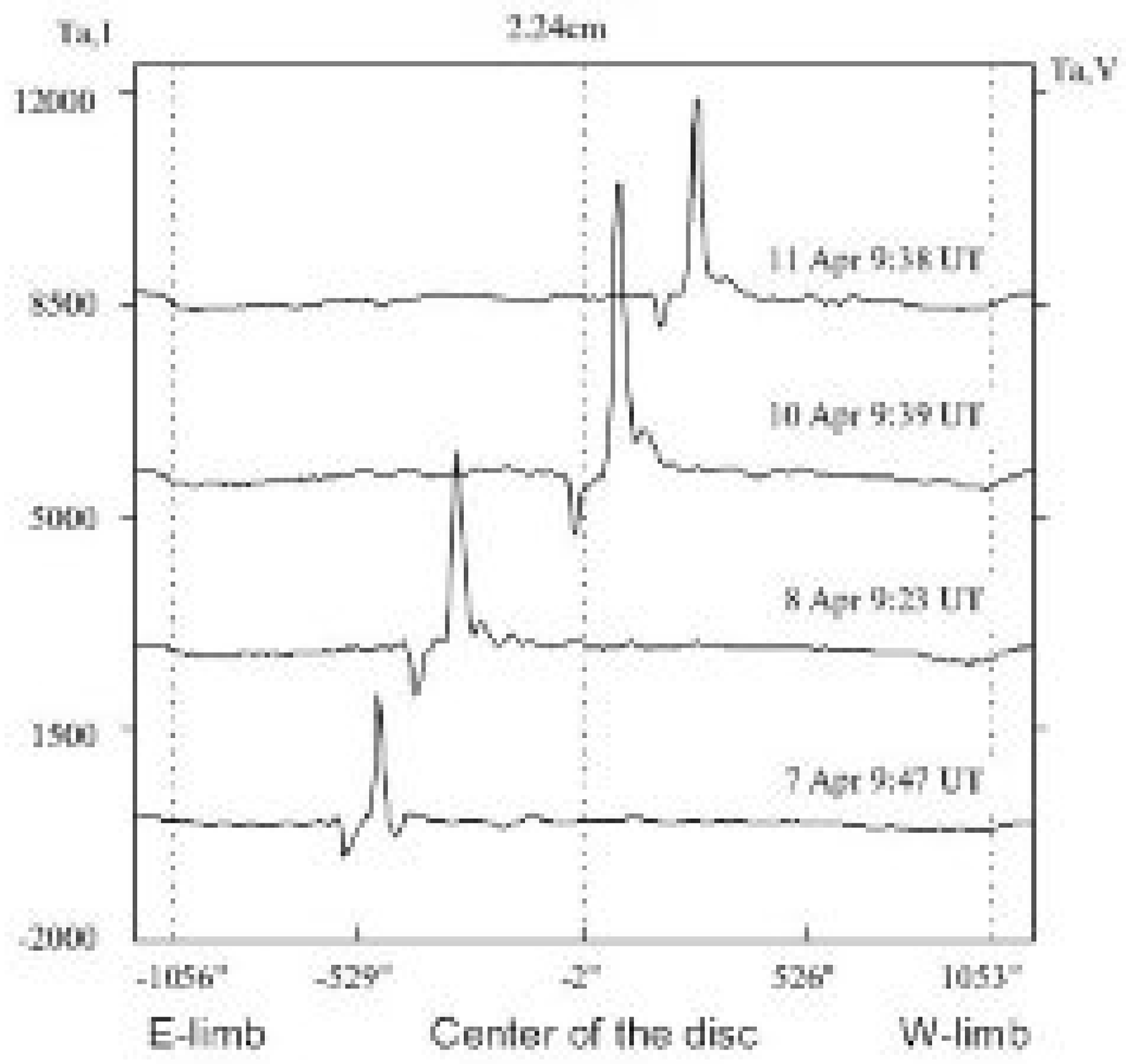}
\hfill
\includegraphics[totalheight=6cm]{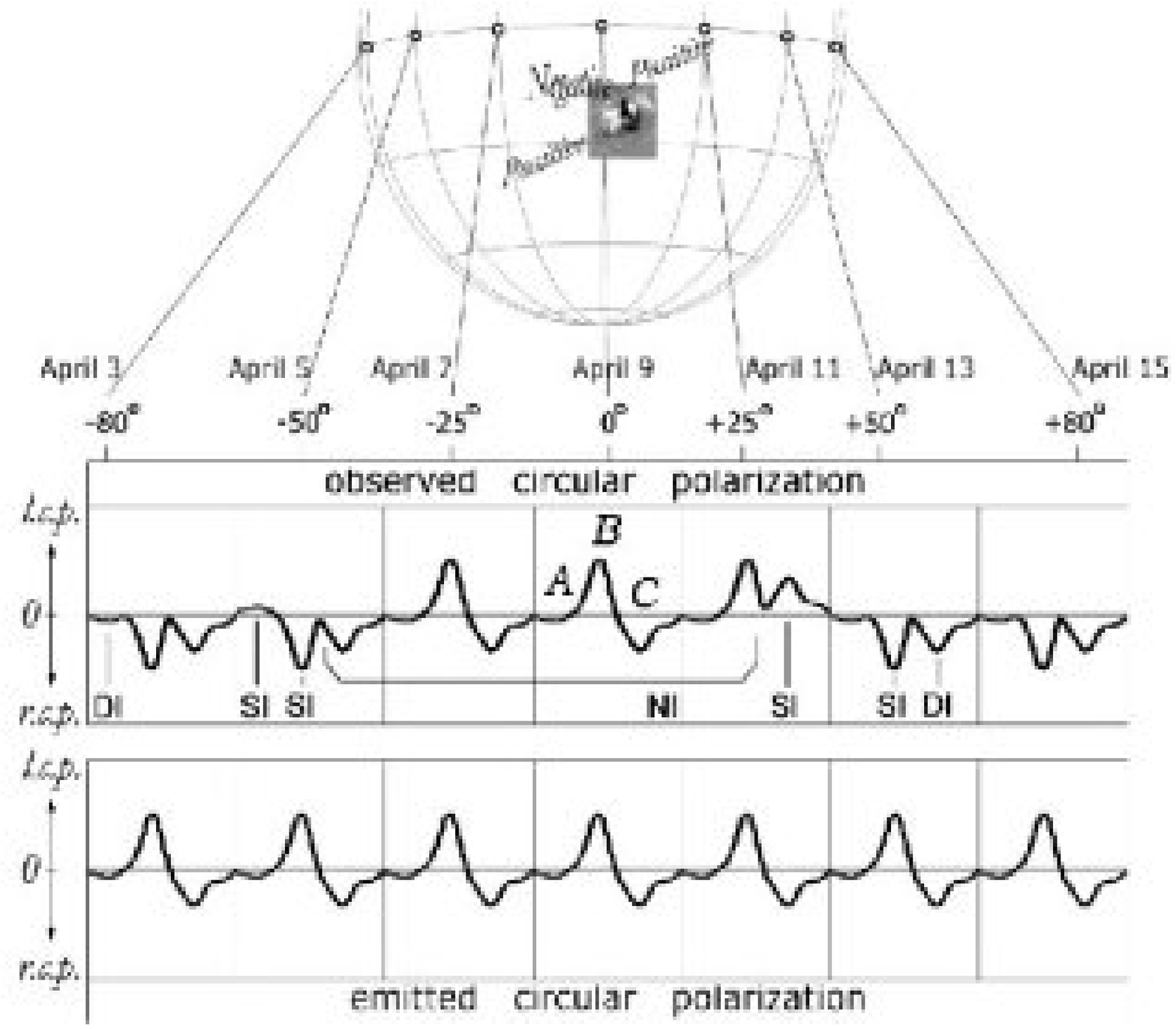}
\parbox[t]{0.47\textwidth}{\caption{The AR~9415 polarization emission scans
at wavelength 2.25cm made on RATAN=600 during
April~7~-~11,~2001, which show R-L-R structure along solar
longitude.}} \label{fig12} \hfill
\parbox[t]{0.47\textwidth}{\caption{ A sketch of polarization changes
in the AR 9415 produced by two coronal regions of
QT-propagation. The polarization inversions in 3 distinct
microwave sources A, B, and C with longitudinal displacement
from central meridian are illustrated by the evolution of V scan
at 2.24cm. The sources found to be of R-L-R sense of circular
polarization while at the center of the solar disk. The A, B,
and C sources are associated with N-S-N magnetic polarities of
the underlying dominant sunspots. The number of polarization
inversions in a distinct microwave source within the "radio
scan" is marked with SI for the singular inversion, with DI for
the double inversion, and with NI for no inversion. Note that
the polarization inversions appear in increasing number as the
AR approaches the solar limb. }} \label{fig13}
\end{figure}

The AR persists as 2 microwave sources A and B of R-L sense of
circular polarization in the SSRT maps (Figure~\ref{fig8}) and
in RATAN-600 scans at the wavelengths longer 5cm on April 7 -
11. The evolution of V scans at the long cm wavelengths clearly
demonstrates the well-known regularities of polarization
inversion \cite{peter74}: the sense of polarization of the most
limb ward source is inverted starting from the long wavelengths
as the AR approaching either solar limb. Thus for the limb ward
source B the first inversion started on April 12 at the
wavelengths longer 6cm turning to the shorter wavelengths on
April 13 (note R-R polarization structure at 5.2 on April 13;
see Figure~\ref{fig8}). The next (second) inversion in the
source B occurred at the wavelengths longer than 6cm on April 14
- 15 (R-L structure). So far as a limb ward microwave source in
western solar hemisphere is concerned. The source A (limb ward
while in eastern hemisphere) is found to invert at the
wavelengths longer than 6cm on April 5 - 6. The inversion turned
to the wavelengths shorter than 5cm on April~6~-~7 (L-L
structure). The next(second, if counting from the central
meridian) inversion in the source A occurred closer to eastern
limb on April 3 - 4, resulted in R-R polarization structure of
the AR at the wavelengths shorter than 5.2cm. As for the longer
wavelengths, the second inversion in the source A has completed
up to April 3. The wavelength-dependent character of
polarization inversion is illustrated by Alissandrakis
\cite{aliss99} with the help of a QT-surface. The near-vertical
QT-surface (where the propagation angle is equal to $90^{o}$)
separates two opposite magnetic fluxes in an AR \cite{aliss96}
and varies its inclination with the longitude. When the source
is near the solar disk center, the microwaves ross QT-surface
higher up in the corona, where the magnetic field is weak and
the resulting polarization is the same as in front of the
QT-surface. As the source moves towards the limb microwaves
cross QT-surface lower in the corona, where the magnetic field
is stronger and the sign of resulting circular polarization
reverses at long wavelengths \cite{zhelez70}. The closer an AR
to solar limb is, the shorter is an operational wavelength
required to detect the polarization inversion. At least two
QT-surfaces are expected to separate N-S-N magnetic fluxes of
the AR 9415. Thus, the source A has a chance to show the double
polarization inversion while on the solar disk
(Figure~\ref{fig13}). During the first inversion, that is, on
April 5 - 7 the AR photospheric fields are stable to the extent
that the convolution with the RATAN-600 beam pattern looks like
the radio scan in Stokes V at the short wavelength of 2.24cm
(Figure~\ref{fig6}). The stable (in a sense of polarity in the
AR over the period of April 5 -12) photospheric magnetic fields
give further evidence that the polarization radio inversion is
caused by the propagation effects in the solar corona. In
practice, the polarization inversion is easy observable and sure
if the underlying microwave source is well resolved. Actually
the limb ward microwave source was the first to invert the sense
of polarization if watched from the solar disk center to either
limb. It was the source A in the eastern hemisphere and C in the
western. This source is found inverted twice at the longitudinal
displacement of $50^{o} - 80^{o}$ from the central meridian,
while the source closest to the meridian is found inverted once.
The longitudinal asymmetry of the AR 9415 polarization
inversions in two solar hemispheres results from the difference
in the strength of the coronal fields in leading and following
QTR. This difference is likely caused by both the different
maximums of the photospheric fields and the dissimilar
inclinations of the QT-surfaces admitted by  \cite{kundu84} and
\cite{lee98}. It is worth noting the speedy completion of each
polarization inversion in the NOAA 9415. Indeed, the only radio
observation of April~6 succeeded to catch the moment of opposite
signs of circular polarization within cm wavelength range in the
course of the inversion. (There is no consideration to the
complicated polarization effects of April 12 - 13.) Other
observations with the RATAN-600 (\cite{gelf87}, \cite{ryabov99}
have got records of both signs of polarization within cm
wavelengths for the microwave source involved in the inversion.
In the case of the AR 9415, all the inversions took much less
than a day for completion in microwave range. We propose the
steep gradients of the coronal magnetic fields in AR 9415 as the
reason for the high speed and the abundance of the inversions.

 \subsubsection { Coronal magnetic field estimates}

We can draw on the equation of the magnetic field B in QTR in
relation to the transition wavelength $\lambda$ to estimate the
strengths of the coronal fields in the magnetosphere of AR 9415.
The equation is derived from the parameter of the wave mode
coupling in the QTR \cite{zhelez70}, provided the product of the
electron density and the scale of the magnetic field divergence
in the QTR is equal $10^{18}$cm$^{-2}$ \cite{bezrukov03}:
\begin{center}
$B_{(G)}\simeq180\cdot\lambda_{\tau}^{-\frac{4}{3}}$
\end{center}
As defined above, the transition wavelength is the wavelength of
zero circular polarization as transition between two wavelengths
ranges where opposite signs of circular polarization are to be
observed. Closer examination of the AR 9415 at the eastern solar
limb on April 3 (Figure~\ref{fig1}) leads us to the conclusion
that the microwave source A is risen to the height of
$4\cdot10^{9}$cm above the limb. Note that the near-limb
polarization inversion in the source A occurs at $\lambda=2.5$
cm (transition frequency equals to 12~GHz; see
Figure~\ref{fig1}). This requires as strong coronal fields as
53~G to produce the observed inversion in the QTR at the height
no less than the height of the source ($h>4\cdot10^{9}$cm). The
microwave sources B and C seem to be  overlapped and polarized
in RH sense. On the evidence of the polarization inversion in
the source B at $\lambda=5$cm on April~6 (Figure~\ref{fig8}) one
can estimate the coronal field to be 21 G in the QTR covering
the source B. The most effective determinations of the relevant
QTR heights imply the detailed model analysis \cite{bezrukov03}.
       The rough estimates can be made by the rate of polarization inversion
\cite{gelf87}. For April~6 observations of the source B
($31\arcsec$ in diameter; 1 day interval taken for polarization
inversion; $\lambda =5$cm)  this method yields a height of
$6\cdot10^{9}$cm for the coronal field of 21~G.

\subsection{The effects of darkening and formation frequency
domain of modes coupling}

The active region 9415 went on to the new state starting from
13th of April. From the one hand, this state is characterized by
appearance of the an intermediate frequency domain of o- and e-
Modes Coupling" (FDMC) with low polarization signal and multiple
inversions of polarization, from the other hand, the effects of
darkening emission in intensity channels were observed
(Figure~\ref{fig11}).

The similar phenomenon was observed before the Bastille day
flare at April~14, 2000 \cite{tokh03}. Here we considerate this
phenomenon in detail due to the frequent multiazimuth
observations. On the Figure~\ref{fig9} the multi-frequency scans
of AR9415 are presented for one azimuth. On the right the FDMC
of this AR is shown on the big scale. One can see that the
signal inside of FDMC band is significantly smaller then out of
it. And at the same time the polarization structure in the band
of 2.67cm - 3.21cm is changed abruptly from one wavelength to
another (Figure~\ref{fig9}). We have compared the activity
inside of  FDMC band at wavelength 2.90cm and outside of this
band at 4.32cm. It is seen that in this AR the leap forward
processes (probably nonthermal) at wavelength 2.90cm were
occurred continuously from 7:53~UT to 10:35~UT (see the AR in
the middle of the left on the Figure~\ref{fig10})). On the right
the temporal behavior of this AR at wavelength 4.32cm during the
same interval of time is shown. One can note there is not any
polarization inversion in the AR, but the corresponded
variations of the polarization emission are presented. This
phenomenon was being exist 3 days until the moment of X14 flare
at 13:50~UT on April 15.

 The spectra of intensity and circular polarization for
five days before the flare are presented on the
Figure~\ref{fig11}. In the intensity channels the effect of
darkening became apparent rather clearly from day to day, and
the gradual formation of FDMS band can be traced in the
polarization channels. The interpretation of this phenomenon was
proposed in paper \cite{tokh03} as the propagation of cyclotron
radio emission through the cold filament based on the model
\cite{zlotnik01}. However, this model does not explain both the
abrupt changes of polarization structures from wave to wave and
the temporal changes (see Figures~\ref{fig9} and \ref{fig10}).
Thus, the analysis of AR 9415 observations revealed the
diversity of polarization characteristics taking place during
early preparatory stage, during the pre-erupting stage and the
main stage of a strong flare.

\section{Conclusions}

The comprehensive investigation of the flare productive active
region NOAA AR 9415 have revealed that polarized emission in
broad range of wavelengths represent a complicated structure of
the active region's magnetosphere. The presence of multiple
polarization inversions could be caused both by a flare
productive nature of the active region and as well as the
geometry of propagation of microwaves through quasi transverse
magnetic fields. To evaluate correctly the magnetic field
parameters in broad range of coronal heights from 1000 km to
some 100000 km it is necessary to perform a spectral analysis of
the emission in a broad wavelength range in combination with
two-dimensional mapping at single waves and with satellite data
on optical and X-ray bands. In the paper for the first time the
complete evolution of QT-effect  over a full time of the AR
passage along the solar disk was studied at several wavelengths
in simultaneous observations.

The polarization inversion is intrinsical feature of
flare-productive active region on the preparatory, pre-eruptive
and eruptive stages. This inversion and other effects (such as
multiple polarization inversions on frequency and on time, the
effect of  microwave darkening before a strong flare, short wave
polarization increasing etc) are sharply  revealed both in
circular polarization emission spectrum and its temporal
changes.
  The new polarized emission phenomenon discovered require
adequate model investigation and appropriate observations with
high parameters in sensitivity, space and temporal resolution.
The high sensitivity of radio waves to magnetic fields could be
used for understanding of active processes going on at different
heights of the solar atmosphere.

\section{Acknowledgments}

The authors thank INTAS (grant 0181 and 0543) which allowed to
join the efforts of different scientific groups in this work.


\end{document}